\documentclass[preprint,onecolumn,preprintnumbers,amsmath,amssymb,floatfix]{revtex4-1}
\usepackage{graphicx}
\usepackage{ gensymb }
\usepackage{soul} % \st{text} for strike out, does not work with \refs inside though
\usepackage[normalem]{ulem}

\usepackage{color}
\newcommand{\comments}[1]{}
\newcommand{\aprox}{$\sim$} 

\begin{document}

\title{Noise in electromigrated nanojunctions}

\author{P. J. Wheeler$^{1}$, Ruoyu Chen$^{1}$, D. Natelson$^{1, 2}$}

\affiliation{$^{1}$ Department of Physics and Astronomy, Rice University, 6100 Main St., Houston, TX 77005}
\affiliation{$^{2}$ Department of Electrical and Computer Engineering, Rice University, 6100 Main St,.Houston, TX 77005}

\date{\today}

\begin{abstract}

Noise measurements are a probe beyond simple electronic transport that can reveal additional information about electronic correlations and inelastic processes.   Here we report noise measurements in individual electromigrated nanojunctions, examining the evolution from the many channel regime to the tunneling regime, using a radio frequency technique.  While we generally observe the dependence of noise on bias expected for shot noise, in approximately 12\% of junction configurations we find discrete changes in the bias dependence at threshold values of the bias, consistent with electronic excitation of local vibrational modes.  Moreover, with some regularity we find significant mesoscopic variation in the magnitude of the noise in particular junctions even with small changes in the accompanying conductance.  In another $\sim$17\% of junctions we  observe pronounced asymmetries in the inferred noise magnitude as a function of bias polarity, suggesting that investigators should be concerned about current-driven ionic motion in the electrodes even at biases well below those used for deliberate electromigration.	
 \end{abstract}

%--------------------------------------------------------------------------------------------------------------------------------

%\pacs{}
\maketitle

Current noise, the fluctuations about the mean rate of charge flow, can reveal much information about the physics at work in nanoscale systems.  In the absence of an applied bias, Johnson-Nyquist noise\cite{Johnson:1928,Nyquist:1928} provides a means of inferring the electronic temperature.  In the presence of a bias, additional excess noise results.  Dynamical ionic degrees of freedom that produce conductance fluctuations as a function of time lead to a contribution quadratic in the bias current\cite{Weissman:1988}.  In the limit of many fluctuators, these fluctuations tend to have a $1/f$ frequency dependence, and are also known as ``flicker'' noise.  Shot noise\cite{Blanter:2000}, the intrinsic excess noise due to the granularity of charge, is detectable in nanoscale systems where electron-phonon scattering processes do not greatly perturb distributions of electrons among particular quantum channels of transmission probability $\tau_{i}$.     

In the limiting case of a classical vacuum diode, charge carriers provide some average current, $I$, from a source to a drain electrode, with some mean traversal rate of electrons, but with the carrier motions otherwise uncorrelated in time.   The electronic current is then governed by Poisson statistics, and fluctuations in the arrival times of the carriers lead\cite{Schottky:1918} to a current noise per unit bandwidth proportional to the average current, $S_{I} = 2 e I$, where $e$ is the magnitude electronic charge.   In general, $S_{I} = F \cdot 2 e I$, where $F$ is defined as the multiplicative Fano factor.
 $F$ may differ from one due to electronic correlations \cite{dePicciotto:1997,Saminadayar:1997}, limited numbers of quantum channels for conduction \cite{Reznikov:1995,Kumar:1996,vandenBrom:1999,Beenakker:2003}, and electron-phonon scattering \cite{Blanter:2000,Mitra:2004,Chen:2005,Koch:2005,Koch:2006,Komnik:2009,Avriller:2009,Haupt:2009,Novotny:2011}.  The full expression for current noise at finite temperature $T$ without inelastic effects is \cite{Blanter:2000}
\begin{equation}
    S_{I} = 4 k_{\mathrm{B}}T G_{0} \sum_{i}\tau_{i}^{2}  +  2 e V G_{0} \coth \frac{e V}{2 k_{\mathrm{B}}T}   \sum_{i} \tau_{i}(1-\tau_{i}),
  \label{eq:shotnoise}
\end{equation}
where $k_{\mathrm{B}}$ is Boltzmann's constant, $V$ is the voltage drop across the junction, and $G_{0}\equiv 2 e^2/h$ is the quantum of conductance.  Temperature appears here entirely because of the thermal broadening of the Fermi-Dirac distributions in the source and drain, and the $\tau_{i}$ are assumed to be temperature-independent and energy-independent on the scale of the bias $eV$.  Likewise, it is assumed that any heating of the electronic distributions due to bias takes place far from the junction, so that $T$ is unaffected by $V$.  At zero temperature the Fano factor reduces to $F = (\sum_{i}\tau_{i}(1-\tau_{i}))/(\sum_{i} \tau_{i})$, showing suppression of the shot noise for fully transmitting channels, as has been verified in semiconductor point contacts and atomic scale metal junctions \cite{vandenBrom:1999,Wheeler:2010,Chen:2012}.  This suppression survives at finite temperatures.

If local bosonic modes such as vibrations are present, there are theoretical predictions\cite{Mitra:2004,Chen:2005,Koch:2005,Koch:2006,Komnik:2009,Avriller:2009,Haupt:2009,Novotny:2011} that $F$ can be modified from this value.  The particular modification depends upon the details of the electron-vibrational coupling and the coupling of that local mode to bulk phonons.  

In this paper we present measurements of excess noise in lithographically created, electromigrated gold nanojunctions at liquid nitrogen (77~K) temperatures, as junctions are migrated from the diffusive regime into a tunneling configuration. While we generally find excess noise compatible with Eq. ~(\ref{eq:shotnoise}), in some junctions we observe asymmetry in measured noise between the positive and negative bias polarities.  In some junctions we also find that atomic-scale configurational changes, detected through their minor impact on the source-drain conductance, can have considerable effects in the measured noise.   In select junctions, we observe a significantly nonlinear dependence of noise as a function of bias, exceeding that expected from shot noise, suggesting that flicker noise plays a significant role in some junctions' characteristics.  We also observe changes in the apparent Fano factor as a function of bias in some junctions, interpreted in mechanical break junction experiments\cite{Kumar:2012} as the result of inelastic processes involving local vibrational modes.   The totality of these results indicates that inelastic electronic coupling to ionic degrees of freedom can be of great importance in these nanoscale junctions, while depending in detail on the atomic-scale junction geometry and disorder.  

\begin{figure}[h!]
\includegraphics[clip, width=8.5cm]{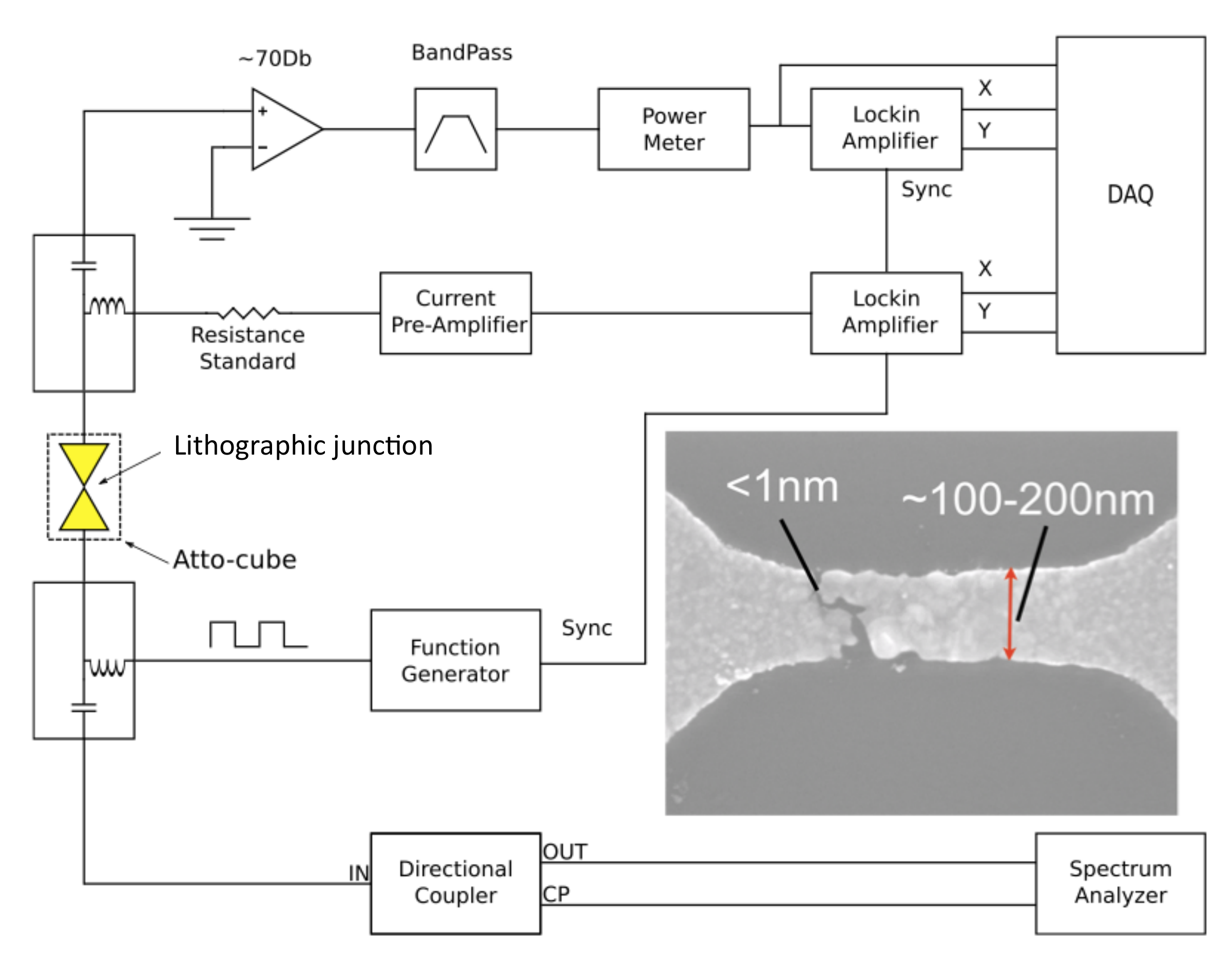}
\caption{Schematic of measurement apparatus. The dashed line indicates the boundary of the $^{4}$He cryostat and enclosure.  Inset:  An electron micrograph of a typical electromigrated junction.}
\label{fig1} 
\end{figure}

Figure~\ref{fig1} shows a schematic of the experimental setup. Our gold bowtie structures are patterned on top of an oxidized Si wafer using electron beam lithography.  On both sides of the 100-200 nm wide constriction, the structure flares out to larger pads, where contact with our electrical probes is made. After development 1 nm Ti and 15 nm Au are evaporated onto the sample followed by lift-off  off with acetone.   The devices are cleaned by oxygen plasma for ~1 minute immediately prior to insertion into the sample insert of a cryostat with four independently positionable probes.  This sample space is evacuated at room temperature and back-filled with 10 mB of He exchange gas.  The sample is then cooled to 77 K for the duration of these experiments.  

Once the sample is cooled to the operating temperature, individual probes are put into electrical contact with the pads. Initial resistance of contacts plus device is \aprox 50 ohms. The devices are electromigrated by applying a ramping bias from 0 to between 0.3 and 0.9 V with a computer-controlled feedback process based on junction resistance determining the maximum ramp voltage. Once a desired dc resistance level is reached, electromigration is halted, allowing the simultaneous measurement of current and noise power as a function of voltage applied across the junction.  We have examined a total of 16 distinct devices in a total of 168 electromigrated configurations.  

A device is contacted within the cryostat by high frequency-compatible, moveable probes.  These probes include proximate ground planes to minimize high frequency insertion losses at the probe tip-sample junction.  The probes are connected to flexible minicoaxial cable (characteristic impedance 50~$\Omega$) within the cryostat, up to SMA connectors at the top of the cryostat.  Transitioning to larger 50~$\Omega$ coaxial cable outside the cryostat, each side of the junction is connected to a bias tee (as shown in Fig.~\ref{fig1}), which plays the crucial role of separating the low and high frequency components of the signal and bias.   

On the low frequency side an offset square wave at a frequency between 2 and 15 kHz (typically 10 kHz for most junction configurations) is sourced so that $V_{\mathrm{min}} \equiv 0$ volts and $V_{\mathrm{max}} \equiv V_{\mathrm{DC}}$ where is the intended driving voltage across the junction.  The square wave is supplied by a SR345 function generator, and the bias tees have been chosen so that distortion of this square wave is minimal.  Varying the applied bias between zero and $V_{\mathrm{DC}}$ enables a SR830 digital lock-in amplifier to measure the change in voltage produced by the power meter that is due to the applied bias.   Combined with the $V=0$ power meter output, the lock-in measurement allows us to determine the change in junction noise due to the junction being driven out of equilibrium.  
$V_{\mathrm{DC}}$ is varied on the timescale of tens of seconds to allow the acquisition of a current-voltage characteristic for the junction.  The low frequency input impedance on the Keithley 428 preamplifier is always more then two orders of magnitude smaller then the resistance of the junction, so virtually all of the sourced voltage is dropped across the junction.  The output of the current preamplifier is measured with a second SR830 lock-in amplifier synchronized to the square wave to find the conductance at each $V_{\mathrm{DC}}$ value.  The high speed X and Y outputs of both lock-in amplifiers are recorded by a National Instruments, NI USB 6259-BNC DAQ.

The efficacy of the excess noise measurement depends on the high frequency impedance response of the circuit.  Both sides of the junction are contacted within the cryostat by high frequency-compatible probes attached to 50~$\Omega$ characteristic impedance coaxial cable.   As shown in Fig.~\ref{fig1}, one high frequency port of the bias tee is terminated with a high frequency 50~$\Omega$ impedance, or connected to a high frequency spectrum analyzer (also with nominal input impedance of 50~$\Omega$). The other high frequency port leads to a series of room temperature rf amplifiers with $\sim$~70 dB of gain,  a bandpass filter that limits the radio frequency range of interest to between 200-600 MHz, and then a logarithmic power detector.   Additional filters are optionally included after the first amplifier in the chain.  Because the lock-in measurement provides the change in power detector voltage due only to the applied (low frequency) bias, much extraneous background noise and bias-independent Johnson-Nyquist noise are thus filtered out.  A more in-depth discussion of this approach can be found in Ref.~\cite{Wheeler:2010}.  

While the system shown in Fig.~\ref{fig1} produces an output proportional to the rf noise produced by the biased junction, calibration of the noise power in absolute units (A$^{2}$/Hz of current noise across the device) is challenging. The junction itself does not have a real 50~$\Omega$ impedance across the full bandwidth of the measurement system, while the remainder of the high frequency measurement circuit is nominally well behaved.  This impedance mismatch implies that some fraction of the RF noise generated by the device is reflected at the device-probe junctions, leading to inefficient out-coupling of the RF noise power.  We estimate this loss for each junction configuration using a reflection measurement, prior to the noise measurement.  Once a junction configuration is prepared through electromigration, a RF spectrum analyzer is used to measure the frequency-dependent reflection coefficient of the device (and measurement wiring) across the filter-defined bandwidth, as described previously in the supplemental material of Ref.~\cite{Wheeler:2010}.   By combining this measurement with a gain-bandwidth product measurement of the amplifier-filter chain, we correlate the voltages measured by the power meter to the current noise power spectral density produced by the junction.  We note that the essential observations reported in this paper do not depend on this calibration procedure or the absolute magnitude of the measured noise.

\begin{figure}[h!]
\includegraphics[clip, width=8.5cm]{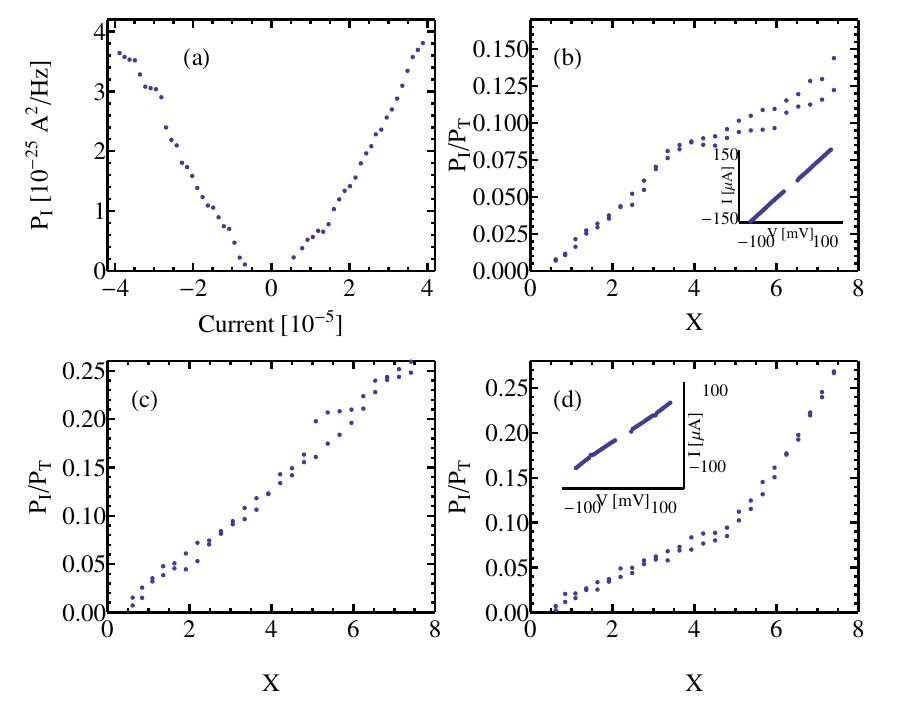}
\caption{Bias dependence of the inferred noise in particular junctions.  (a) Inferred excess noise as a function of current in a junction with zero-bias resistance of 3~k$\Omega$, showing intrinsic nonlinearity.  (c)  A scaled plot of normalized excess noise as a function of scaled bias for the junction in (a).  The linearity of this plot shows that the nonlinearity in (a) originates from the expected functional form of finite-temperature shot noise, as in Eq.~(\ref{eq:shotnoise}).    (b), (d) Examples of discrete changes in Fano factor at particular threshold voltage biases of tens of mV.  Such changes in slope are consistent with previously reported observations\protect{\cite{Kumar:2012}} of inelastic electronic interactions with local vibrational modes, though in this case the junctions involve many conductance channels.  The junction in (b) has a resistance of 600~$\Omega$, while that in (d) has a resistance of $1.1$~k$\Omega$.  Insets show the relevant simultaneously acquired $I$ as a function of $V_{\mathrm{DC}}$ data.  These data do not show perceptible kinks at the voltages relevant for the changes in Fano factor, though this is not particularly surprising given that lock-in techniques are generally necessary to resolve such features.}
\label{fig2} 
\end{figure}

Figure \ref{fig2} shows raw, averaged current noise as a function of DC current for a particular junction electromigrated to a DC resistance of approximately 3~k$\Omega$, so that conduction is expected to proceed through a small number of quantum channels.  Some nonlinearity is present in such a plot, and it is necessary to determine whether this is originates from finite temperature effects or other noise processes.  When comparing the bias-dependence of the noise with the simple finite temperature expectations of Eq.~(\ref{eq:shotnoise}), we follow the approach of Kumar~\textit{et al.}\cite{Kumar:2012} by scaling the data for plotting purposes.  The change in current noise due to the bias, $P_{I} \equiv S_{I}(V_{\mathrm{DC}}) - S_{I}(0)$, inferred from the lock-in measurements of the power meter, is normalized by the expected Johnson-Nyquist noise, $P_{T} \equiv 4 k_{\mathrm{B}}T G$, and plotted as a function of a scaled bias coordinate:
\begin{equation}
    X(V_{\mathrm{DC}}) = \frac{e V_{\mathrm{DC}}}{2 k_{\mathrm{B}}T}  \coth \frac{e V_{\mathrm{DC}}}{2 k_{\mathrm{B}}T}  - 1.
  \label{eq:scaledx}
\end{equation}
In these scaled plots, normalized noise in accord with Eq.~(\ref{eq:shotnoise}) should be linear in $X(V_{\mathrm{DC}})$ with a slope given by the zero-temperature Fano factor and an intercept of zero at $X=0$, with identical response for positive or negative polarity of $V_{\mathrm{DC}}$.    

At least 86 junction configurations examined show simple linear noise data consistent with these expectations (a number of additional junction configurations had insufficient data collection to guarantee reproducible, stable linear response at both polarities of bias).  An example of this linear dependence is seen in Fig.~\ref{fig2}(c).  Assuming that the calibration procedure described above is absolutely correct, the Fano factor for this device is approximately 0.036.  The simplest way of reconciling this with the restriction of a small number of contributing channels is that the calibration procedure is not precise and that conduction in this particular junction is dominated by four nearly fully transmitting channels.  The important point for this and similar junction configurations is that the bias dependence of the noise is functionally identical to the expectations of Eq.~(\ref{eq:shotnoise}) assuming a constant Fano factor over the applied bias range.

In contrast, some devices show clear deviations from expected linear dependence of $P_{I}/P_{T}$ on $X$, suggesting either a voltage-dependent Fano factor, or noise processes not encompassed by Eq. \ref{eq:shotnoise}.   Examples of such deviation are are kinks in the noise as a function of bias when $V_{\mathrm{DC}}$ is in the tens of millivolts range, as seen in Fig. \ref{fig2} (b), (d).  Assuming that the noise is the excess noise of Eq. \ref{eq:shotnoise}, the sudden change of slope  at a particular bias indicates a change in Fano factor.   Similar changes in slope have been observed in 20 junction configurations, sometimes with slopes increasing above a critical voltage (Fig.~\ref{fig2}(d)), sometimes with slopes decreasing above a critical voltage (Fig.~\ref{fig2}(b)).  In such junctions, these kinks are repeatable and stable over many bias sweeps.  In terms of voltage across the junction, these kinks appear at an average of 52 mV bias, with a standard deviation of 14 mV.  This bias level is approximately an order of magnitude below the bias range where electromigration is normally observed.   We note that there is no obvious corresponding non-linearity observed in the current-voltage characteristics, though the measurement technique used here (measuring $I$ directly at each value of $V_{\mathrm{DC}}$) precludes easy simultaneous measurement of the higher derivatives of $I$ vs. $V_{\mathrm{DC}}$, traditionally performed using lock-in methods.    The sensitivity of device properties to atomic-scale details of junction configurations has made it very difficult to change back and forth between the standard lock-in approach for measuring $dI/dV$ and $d^{2}I/dV^{2}$ and the noise measurement approach without affecting the device under test.  Attempts to improve the situation are ongoing.

Voltage-thresholded changes in the excess noise have been described in a large number of models that describe electronic transport with a small number of transmitting channels in the presence of electronic coupling to a local vibrational mode \cite{Mitra:2004,Chen:2005,Komnik:2009,Avriller:2009,Haupt:2009,Novotny:2011}.  For example, recently, Kumar \textit{et al.}observe qualitatively similar kinks close to $G = 1 G_{0}$ in atomic-scale gold junctions (dominated by a single nearly-fully transmitting channel) prepared through a mechanical breakjunction approach\cite{Kumar:2012}.  Those kinks are clustered in bias near 17 mV, an energy scale comparable to an optical phonon in Au, are attributed to electrons coupling to this mode in the atomic-scale nanowires.   Moreover, those authors find a systematic relationship between the sign of the change in Fano factor, $\delta F$, when $V_{\mathrm{DC}}$ crosses the kink, and the total junction conductance.  In contrast, our measurements show qualitatively similar kinds in junctions traversed by as many as 20 channels (based on the total conductance of the junction, which is still limited by the constriction rather than the leads), with no obvious correlation between total conductance and sign or magnitude of $\delta F$.  This implies that such electron-vibrational modifications to the excess noise are still relevant even in many-channel junctions, with the sign and magnitude of $\delta F$ being set by the transmittance of the particular channel that happens to be strongly coupled to the local vibrational mode.   While the variation of the kink voltage scale in our junctions is significant, the rough magnitude suggests that the inelastic mode could be due to an adsorbate such as molecular hydrogen\cite{Djukic:2005} rather than optical phonons of the metal itself.

% Along with recent theoretical predictions that vibration modes can modifying the Fanno factor %\cite{Mitra:2004,Chen:2005,Koch:2005,Koch:2006,Komnik:2009,Avriller:2009,Haupt:2009,Novotny:2011}(\textcolor{blue}%{Should this just be chen, \cite{Chen:2005, Novotny:2011} since they are the only ones to talk about point contacts?}) %additional recent experimental results \cite{kumar:2012}, show evidence of just that. The example kinks shown in Fig. %\ref{fig2} (c, d) are potentially due to electron-phonon interaction occur above some critical voltage ~50-70mV. While %this voltage range is often observed our data set does not currently is not large enough to proper statistics and an area %for future consideration. Unlike what you would expect form $1/4$ noise the noise does not scale quadratically with %voltage.  Rather the Fanno factor takes on a constant value both above and below the threshold voltage.  Nor does it %scale as $\sqrt{V}$ as would be expected if the local thermal energy was greater then the smallest vibrational mode %energy \cite{Chen:2005}. While this is not conclusive is indicative and is some overlap with \cite{kumar:2012} .

\begin{figure}[h!]
\includegraphics[clip, width=8.5cm]{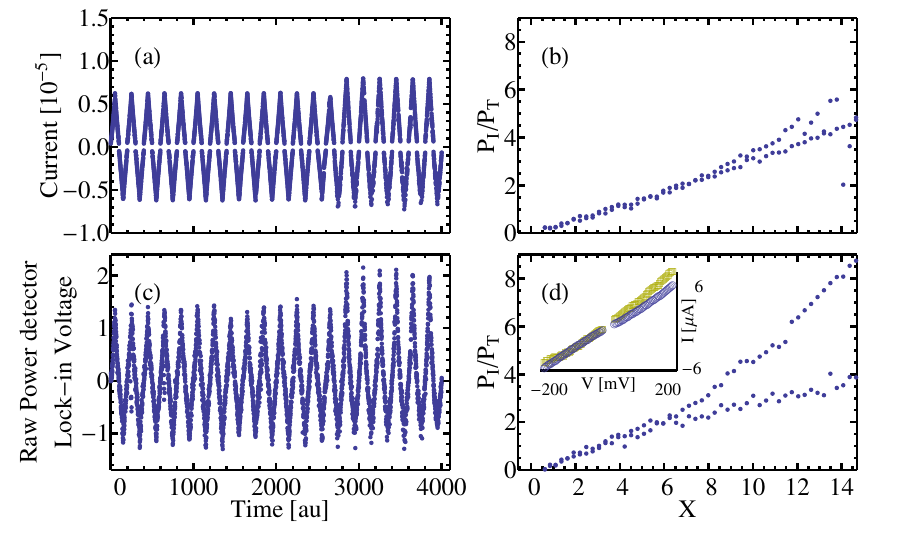}
\caption{Noise asymmetry and mesoscopic sensitivity.  (a) Current as a function of time as  $V_{\mathrm{DC}}$ is swept in a triangle wave between -200~mV and +200~mV, showing a stochastic transition between two junction configurations with similar conductances.  Zero-bias resistance of the junction is 34~k$\Omega$.  (b)  The averaged, normalized excess noise as a function of scaled bias for the first configuration, showing two distinct branches, the upper branch with higher maximum noise corresponding to positive bias polarity.  (c) The raw lock-in measurement of the RF power detector as a function of time, showing clearly that the second junction configuration exhibits noise much more asymmetric in bias than the current itself in (a).  (d) Normalized excess noise as a function of scaled bias in the second configuration, with the upper branch again corresponding to positive bias polarity.  The inset to panel (d) shows $I$ as a function of $V_{\mathrm{DC}}$ before (open symbols) and after (filled symbols) the change in junction configuration.  The change in noise asymmetry is considerably more dramatic.}
\label{fig3} 
\end{figure}

In addition to these kinks, we see additional evidence for deviations in the noise from the simple expectations of Eq.~(\ref{eq:shotnoise}) in some fraction of  junctions.   Figure~\ref{fig3} shows a clear example demonstrating two particular features.  Fig.~\ref{fig3}(a) shows the current (strictly speaking, the change in $I$ due to the square wave from $0$ to $V_{\mathrm{DC}}$)  as a function of time (au) as $V_{\mathrm{DC}}$ is varied up and down through both polarities up to a maximum magnitude of 100 mV.   At a time index of approximately 2600, there is a clear configurational change in the junction, as seen by the obvious change in $I$ as a function of $V_{\mathrm{DC}}$.   Fig.~\ref{fig3}(b) shows the corresponding $P_{I}/P_{T}$ vs. $X$ averaged over the bias sweeps in the initial junction configuration, while Fig.~\ref{fig3}(d) shows the noise \textit{after} the configurational change.  In both configurations there are two distinct branches apparent, indicating that there is an asymmetry in the noise response between the different bias polarities, with positive bias corresponding to higher inferred bias-driven noise.  In 28 junction configurations we observe some distinct asymmetry in $P_{I}/P_{T}$ vs. $X$ as a function of the polarity of $V_{\mathrm{DC}}$.  This asymmetry is truly intrinsic to the particular junction configurations, confirmed by swapping the biasing circuitry and current amplifier between the source and drain electrodes.  In most of these configurations, as here, the inferred $P_{I}/P_{T}$ is smoothly nonlinear (no indication of sharp kinks) in $X$ in at least one polarity.   Note in this case that at negative bias polarity, the linearity of $P_{I}/P_{T}$ with $X$ is excellent, consistent with Eq.~(\ref{eq:shotnoise}) and a Fano factor of approximately 0.7.   At positive bias polarity, there is a smooth, superlinear in $X$, contribution to the noise.

We find that asymmetries of this type (deviation from linear shot noise at one bias polarity) can also arise stochastically and irreversibly due to atomic-scale junction rearrangement at comparatively large biases ($>$ 150 mV), particularly in junctions with conductances $< 0.7~G_0$.  These changes in noise asymmetry can take place even when the accompanying change in conductance is comparatively minimal (see inset to Fig.~\ref{fig3}(d)).   The configurational change in Fig.~\ref{fig3} is a spontaneous example.  After the atomic rearrangement, the zero-bias conductance increases by approximately 10\%, with a slightly higher differential conductance at positive polarity.  Figure~\ref{fig3}(c) shows the corresponding time variation of the raw lock-in amplifier reading of the RF power meter.  After this modest change in conductance, there is now a much larger asymmetry in the inferred noise, with $P_{I}/P_{T}$ at maximum positive bias exceeding that at negative bias by more than a factor of two.  

A natural source of such a nonlinearity would be flicker noise, with time-dependent fluctuations of the junction conductance leading to a fluctuating current; however, ordinary time-dependent conductance fluctuations should depend only on the presence and properties of fluctuators in the junction region, and do not by themselves imply an asymmetry with bias.   There has been comparatively little discussion of noise processes dependent on current polarity in the literature.  Experimentally, shot noise measurements in carbon nanotubes have shown asymmetry as a function of bias \cite{Wu:2007}, ascribed to the same interference between conducting channels that gives rise to Fabry-Perot resonances in the conductance.  Atomic-scale asymmetries in junctions have also been predicted to lead to a possible dependence of the noise on bias polarity\cite{Yao:2006}, and this would be sensitive to individual atomic positions.  However, given the significant nonlinearity in the noise we observe in these asymmetric examples, we believe it more likely that some form of current-driven ionic motion tied to a flicker/conductance fluctuation mechanism is at work in our case.  Time-dependent conductance fluctuations in metal systems are believed to arise from dynamical motion of atoms or groups of atoms \cite{Feng:1986,Weissman:1988}.   

We suggest that a likely explanation for our observed asymmetry in this contribution to the noise is local heating of ionic degrees of freedom that depends on the direction of the current.  Under bias, the electronic distribution functions in the contacts near the junction are driven out of thermal equilibrium, with a population of "hot" electrons injected into the positively biased contact.  Such an electronic hot spot can lead to an asymmetric heating of local ionic degrees of freedom\cite{Dagosta:2006}, as has recently been observed experimentally \cite{Tsutsui:2012}.   An alternative mechanism could be the bias-driven runaway pumping of a locally unstable ionic vibrational mode in one of the contacts\cite{Lu:2010}.  In both cases, the idea is that a localized ionic degree of freedom present in one electrode but not the other interacts with the local electronic distribution within an electron-electron inelastic scattering length of the junction.  The application of a bias drives the electronic distribution out of thermal equilibrium, the particular form of the local electronic distribution (and thus the effect on the ionic degree of freedom) depending on the direction of current flow.    Further investigations of junctions exhibiting these asymmetries, including their detailed bias dependence, are warranted, in the hopes of determining more information about the microscopic structural differences responsible.  

We have used high frequency methods to examine noise in individual electromigrated nanojunctions.  Many junctions show a bias dependence of the noise response consistent with conventional shot noise (Eq.~(\ref{eq:shotnoise})).  However, we find two significant deviations from simple shot noise in a significant population of junction configurations, suggestive of current-driven inelastic processes involving ionic degrees of freedom.   Even in junctions with comparatively large numbers of quantum channels contributing to the conduction, we observe discrete changes in the Fano factor of the noise at biases in the range of tens of mV.  These are qualitatively similar to results observed\cite{Kumar:2012} in atomic-scale junctions that are ascribed to inelastic processes involving local vibrational modes, suggesting that such processes are likely at work even in these larger junctions.  In some junctions we find an additional contribution to the noise with a bias dependence inconsistent with shot noise, and likely associated with conductance fluctuations due to motion of ionic degrees of freedom.  This additional noise can be asymmetric in bias current polarity, and is clearly sensitive to atomic-scale rearrangements of the junction configuration.  This sensitivity and the asymmetry of this noise with bias current highlight the critical importance of understanding electronic and ionic heating at the nanoscale, both for future scientific investigations (e.g., current noise in molecular junctions) and nanoelectronics technologies.

%--------------------------------------------------------------------------------------------------------------------------------

%[Insert concluding paragraph.]

PJW, RC, and DN acknowledge support from NSF grant DMR-0855607.

\clearpage

\clearpage

%\bibliographystyle{apsrev4-1}
%\bibliography{./shotnoise.bib}

\begin{thebibliography}{29}%
\makeatletter
\providecommand \@ifxundefined [1]{%
 \@ifx{#1\undefined}
}%
\providecommand \@ifnum [1]{%
 \ifnum #1\expandafter \@firstoftwo
 \else \expandafter \@secondoftwo
 \fi
}%
\providecommand \@ifx [1]{%
 \ifx #1\expandafter \@firstoftwo
 \else \expandafter \@secondoftwo
 \fi
}%
\providecommand \natexlab [1]{#1}%
\providecommand \enquote  [1]{``#1''}%
\providecommand \bibnamefont  [1]{#1}%
\providecommand \bibfnamefont [1]{#1}%
\providecommand \citenamefont [1]{#1}%
\providecommand \href@noop [0]{\@secondoftwo}%
\providecommand \href [0]{\begingroup \@sanitize@url \@href}%
\providecommand \@href[1]{\@@startlink{#1}\@@href}%
\providecommand \@@href[1]{\endgroup#1\@@endlink}%
\providecommand \@sanitize@url [0]{\catcode `\\12\catcode `\$12\catcode
  `\&12\catcode `\#12\catcode `\^12\catcode `\_12\catcode `\%12\relax}%
\providecommand \@@startlink[1]{}%
\providecommand \@@endlink[0]{}%
\providecommand \url  [0]{\begingroup\@sanitize@url \@url }%
\providecommand \@url [1]{\endgroup\@href {#1}{\urlprefix }}%
\providecommand \urlprefix  [0]{URL }%
\providecommand \Eprint [0]{\href }%
\providecommand \doibase [0]{http://dx.doi.org/}%
\providecommand \selectlanguage [0]{\@gobble}%
\providecommand \bibinfo  [0]{\@secondoftwo}%
\providecommand \bibfield  [0]{\@secondoftwo}%
\providecommand \translation [1]{[#1]}%
\providecommand \BibitemOpen [0]{}%
\providecommand \bibitemStop [0]{}%
\providecommand \bibitemNoStop [0]{.\EOS\space}%
\providecommand \EOS [0]{\spacefactor3000\relax}%
\providecommand \BibitemShut  [1]{\csname bibitem#1\endcsname}%
\let\auto@bib@innerbib\@empty
%</preamble>
\bibitem [{\citenamefont {Johnson}(1928)}]{Johnson:1928}%
  \BibitemOpen
  \bibfield  {author} {\bibinfo {author} {\bibfnamefont {J.~B.}\ \bibnamefont
  {Johnson}},\ }\href {\doibase 10.1103/PhysRev.32.97} {\bibfield  {journal}
  {\bibinfo  {journal} {Phys. Rev.}\ }\textbf {\bibinfo {volume} {32}},\
  \bibinfo {pages} {97} (\bibinfo {year} {1928})}\BibitemShut {NoStop}%
\bibitem [{\citenamefont {Nyquist}(1928)}]{Nyquist:1928}%
  \BibitemOpen
  \bibfield  {author} {\bibinfo {author} {\bibfnamefont {H.}~\bibnamefont
  {Nyquist}},\ }\href {\doibase 10.1103/PhysRev.32.110} {\bibfield  {journal}
  {\bibinfo  {journal} {Phys. Rev.}\ }\textbf {\bibinfo {volume} {32}},\
  \bibinfo {pages} {110} (\bibinfo {year} {1928})}\BibitemShut {NoStop}%
\bibitem [{\citenamefont {Weissman}(1988)}]{Weissman:1988}%
  \BibitemOpen
  \bibfield  {author} {\bibinfo {author} {\bibfnamefont {M.~B.}\ \bibnamefont
  {Weissman}},\ }\href {\doibase 10.1103/RevModPhys.60.537} {\bibfield
  {journal} {\bibinfo  {journal} {Rev. Mod. Phys.}\ }\textbf {\bibinfo {volume}
  {60}},\ \bibinfo {pages} {537} (\bibinfo {year} {1988})}\BibitemShut
  {NoStop}%
\bibitem [{\citenamefont {Blanter}\ and\ \citenamefont
  {B{\"u}ttiker}(2000)}]{Blanter:2000}%
  \BibitemOpen
  \bibfield  {author} {\bibinfo {author} {\bibfnamefont {Y.}~\bibnamefont
  {Blanter}}\ and\ \bibinfo {author} {\bibfnamefont {M.}~\bibnamefont
  {B{\"u}ttiker}},\ }\href {\doibase 10.1016/S0370-1573(99)00123-4} {\bibfield
  {journal} {\bibinfo  {journal} {Physics Reports}\ }\textbf {\bibinfo {volume}
  {336}},\ \bibinfo {pages} {1 } (\bibinfo {year} {2000})}\BibitemShut
  {NoStop}%
\bibitem [{\citenamefont {Schottky}(1918)}]{Schottky:1918}%
  \BibitemOpen
  \bibfield  {author} {\bibinfo {author} {\bibfnamefont {W.}~\bibnamefont
  {Schottky}},\ }\href@noop {} {\bibfield  {journal} {\bibinfo  {journal} {Ann.
  der Physik}\ }\textbf {\bibinfo {volume} {57}},\ \bibinfo {pages} {541}
  (\bibinfo {year} {1918})}\BibitemShut {NoStop}%
\bibitem [{\citenamefont {de~Picciotto}\ \emph {et~al.}(1997)\citenamefont
  {de~Picciotto}, \citenamefont {Reznikov}, \citenamefont {Heiblum},
  \citenamefont {Umansky}, \citenamefont {Bunin},\ and\ \citenamefont
  {Mahalu}}]{dePicciotto:1997}%
  \BibitemOpen
  \bibfield  {author} {\bibinfo {author} {\bibfnamefont {R.}~\bibnamefont
  {de~Picciotto}}, \bibinfo {author} {\bibfnamefont {M.}~\bibnamefont
  {Reznikov}}, \bibinfo {author} {\bibfnamefont {M.}~\bibnamefont {Heiblum}},
  \bibinfo {author} {\bibfnamefont {V.}~\bibnamefont {Umansky}}, \bibinfo
  {author} {\bibfnamefont {G.}~\bibnamefont {Bunin}}, \ and\ \bibinfo {author}
  {\bibfnamefont {D.}~\bibnamefont {Mahalu}},\ }\href@noop {} {\bibfield
  {journal} {\bibinfo  {journal} {Nature}\ }\textbf {\bibinfo {volume} {389}},\
  \bibinfo {pages} {162} (\bibinfo {year} {1997})}\BibitemShut {NoStop}%
\bibitem [{\citenamefont {Saminadayar}\ \emph {et~al.}(1997)\citenamefont
  {Saminadayar}, \citenamefont {Glattli}, \citenamefont {Jin},\ and\
  \citenamefont {Etienne}}]{Saminadayar:1997}%
  \BibitemOpen
  \bibfield  {author} {\bibinfo {author} {\bibfnamefont {L.}~\bibnamefont
  {Saminadayar}}, \bibinfo {author} {\bibfnamefont {D.~C.}\ \bibnamefont
  {Glattli}}, \bibinfo {author} {\bibfnamefont {Y.}~\bibnamefont {Jin}}, \ and\
  \bibinfo {author} {\bibfnamefont {B.}~\bibnamefont {Etienne}},\ }\href
  {\doibase 10.1103/PhysRevLett.79.2526} {\bibfield  {journal} {\bibinfo
  {journal} {Phys. Rev. Lett.}\ }\textbf {\bibinfo {volume} {79}},\ \bibinfo
  {pages} {2526} (\bibinfo {year} {1997})}\BibitemShut {NoStop}%
\bibitem [{\citenamefont {Reznikov}\ \emph {et~al.}(1995)\citenamefont
  {Reznikov}, \citenamefont {Heiblum}, \citenamefont {Shtrikman},\ and\
  \citenamefont {Mahalu}}]{Reznikov:1995}%
  \BibitemOpen
  \bibfield  {author} {\bibinfo {author} {\bibfnamefont {M.}~\bibnamefont
  {Reznikov}}, \bibinfo {author} {\bibfnamefont {M.}~\bibnamefont {Heiblum}},
  \bibinfo {author} {\bibfnamefont {H.}~\bibnamefont {Shtrikman}}, \ and\
  \bibinfo {author} {\bibfnamefont {D.}~\bibnamefont {Mahalu}},\ }\href
  {\doibase 10.1103/PhysRevLett.75.3340} {\bibfield  {journal} {\bibinfo
  {journal} {Phys. Rev. Lett.}\ }\textbf {\bibinfo {volume} {75}},\ \bibinfo
  {pages} {3340} (\bibinfo {year} {1995})}\BibitemShut {NoStop}%
\bibitem [{\citenamefont {Kumar}\ \emph {et~al.}(1996)\citenamefont {Kumar},
  \citenamefont {Saminadayar}, \citenamefont {Glattli}, \citenamefont {Jin},\
  and\ \citenamefont {Etienne}}]{Kumar:1996}%
  \BibitemOpen
  \bibfield  {author} {\bibinfo {author} {\bibfnamefont {A.}~\bibnamefont
  {Kumar}}, \bibinfo {author} {\bibfnamefont {L.}~\bibnamefont {Saminadayar}},
  \bibinfo {author} {\bibfnamefont {D.~C.}\ \bibnamefont {Glattli}}, \bibinfo
  {author} {\bibfnamefont {Y.}~\bibnamefont {Jin}}, \ and\ \bibinfo {author}
  {\bibfnamefont {B.}~\bibnamefont {Etienne}},\ }\href {\doibase
  10.1103/PhysRevLett.76.2778} {\bibfield  {journal} {\bibinfo  {journal}
  {Phys. Rev. Lett.}\ }\textbf {\bibinfo {volume} {76}},\ \bibinfo {pages}
  {2778} (\bibinfo {year} {1996})}\BibitemShut {NoStop}%
\bibitem [{\citenamefont {van~den Brom}\ and\ \citenamefont {van
  Ruitenbeek}(1999)}]{vandenBrom:1999}%
  \BibitemOpen
  \bibfield  {author} {\bibinfo {author} {\bibfnamefont {H.~E.}\ \bibnamefont
  {van~den Brom}}\ and\ \bibinfo {author} {\bibfnamefont {J.~M.}\ \bibnamefont
  {van Ruitenbeek}},\ }\href {\doibase 10.1103/PhysRevLett.82.1526} {\bibfield
  {journal} {\bibinfo  {journal} {Phys. Rev. Lett.}\ }\textbf {\bibinfo
  {volume} {82}},\ \bibinfo {pages} {1526} (\bibinfo {year}
  {1999})}\BibitemShut {NoStop}%
\bibitem [{\citenamefont {Beenakker}\ and\ \citenamefont
  {Schonenberger}(2003)}]{Beenakker:2003}%
  \BibitemOpen
  \bibfield  {author} {\bibinfo {author} {\bibfnamefont {C.}~\bibnamefont
  {Beenakker}}\ and\ \bibinfo {author} {\bibfnamefont {C.}~\bibnamefont
  {Schonenberger}},\ }\href {\doibase 10.1063/1.1583532} {\bibfield  {journal}
  {\bibinfo  {journal} {Physics Today}\ }\textbf {\bibinfo {volume} {56}},\
  \bibinfo {pages} {37} (\bibinfo {year} {2003})}\BibitemShut {NoStop}%
\bibitem [{\citenamefont {Mitra}\ \emph {et~al.}(2004)\citenamefont {Mitra},
  \citenamefont {Aleiner},\ and\ \citenamefont {Millis}}]{Mitra:2004}%
  \BibitemOpen
  \bibfield  {author} {\bibinfo {author} {\bibfnamefont {A.}~\bibnamefont
  {Mitra}}, \bibinfo {author} {\bibfnamefont {I.}~\bibnamefont {Aleiner}}, \
  and\ \bibinfo {author} {\bibfnamefont {A.~J.}\ \bibnamefont {Millis}},\
  }\href {\doibase 10.1103/PhysRevB.69.245302} {\bibfield  {journal} {\bibinfo
  {journal} {Phys. Rev. B}\ }\textbf {\bibinfo {volume} {69}},\ \bibinfo
  {pages} {245302} (\bibinfo {year} {2004})}\BibitemShut {NoStop}%
\bibitem [{\citenamefont {Chen}\ and\ \citenamefont
  {Di~Ventra}(2005)}]{Chen:2005}%
  \BibitemOpen
  \bibfield  {author} {\bibinfo {author} {\bibfnamefont {Y.-C.}\ \bibnamefont
  {Chen}}\ and\ \bibinfo {author} {\bibfnamefont {M.}~\bibnamefont
  {Di~Ventra}},\ }\href {\doibase 10.1103/PhysRevLett.95.166802} {\bibfield
  {journal} {\bibinfo  {journal} {Phys. Rev. Lett.}\ }\textbf {\bibinfo
  {volume} {95}},\ \bibinfo {pages} {166802} (\bibinfo {year}
  {2005})}\BibitemShut {NoStop}%
\bibitem [{\citenamefont {Koch}\ and\ \citenamefont {von
  Oppen}(2005)}]{Koch:2005}%
  \BibitemOpen
  \bibfield  {author} {\bibinfo {author} {\bibfnamefont {J.}~\bibnamefont
  {Koch}}\ and\ \bibinfo {author} {\bibfnamefont {F.}~\bibnamefont {von
  Oppen}},\ }\href {\doibase 10.1103/PhysRevLett.94.206804} {\bibfield
  {journal} {\bibinfo  {journal} {Phys. Rev. Lett.}\ }\textbf {\bibinfo
  {volume} {94}},\ \bibinfo {pages} {206804} (\bibinfo {year}
  {2005})}\BibitemShut {NoStop}%
\bibitem [{\citenamefont {Koch}\ \emph {et~al.}(2006)\citenamefont {Koch},
  \citenamefont {von Oppen},\ and\ \citenamefont {Andreev}}]{Koch:2006}%
  \BibitemOpen
  \bibfield  {author} {\bibinfo {author} {\bibfnamefont {J.}~\bibnamefont
  {Koch}}, \bibinfo {author} {\bibfnamefont {F.}~\bibnamefont {von Oppen}}, \
  and\ \bibinfo {author} {\bibfnamefont {A.~V.}\ \bibnamefont {Andreev}},\
  }\href {\doibase 10.1103/PhysRevB.74.205438} {\bibfield  {journal} {\bibinfo
  {journal} {Phys. Rev. B}\ }\textbf {\bibinfo {volume} {74}},\ \bibinfo
  {pages} {205438} (\bibinfo {year} {2006})}\BibitemShut {NoStop}%
\bibitem [{\citenamefont {Schmidt}\ and\ \citenamefont
  {Komnik}(2009)}]{Komnik:2009}%
  \BibitemOpen
  \bibfield  {author} {\bibinfo {author} {\bibfnamefont {T.~L.}\ \bibnamefont
  {Schmidt}}\ and\ \bibinfo {author} {\bibfnamefont {A.}~\bibnamefont
  {Komnik}},\ }\href {\doibase 10.1103/PhysRevB.80.041307} {\bibfield
  {journal} {\bibinfo  {journal} {Phys. Rev. B}\ }\textbf {\bibinfo {volume}
  {80}},\ \bibinfo {pages} {041307} (\bibinfo {year} {2009})}\BibitemShut
  {NoStop}%
\bibitem [{\citenamefont {Avriller}\ and\ \citenamefont
  {Levy~Yeyati}(2009)}]{Avriller:2009}%
  \BibitemOpen
  \bibfield  {author} {\bibinfo {author} {\bibfnamefont {R.}~\bibnamefont
  {Avriller}}\ and\ \bibinfo {author} {\bibfnamefont {A.}~\bibnamefont
  {Levy~Yeyati}},\ }\href {\doibase 10.1103/PhysRevB.80.041309} {\bibfield
  {journal} {\bibinfo  {journal} {Phys. Rev. B}\ }\textbf {\bibinfo {volume}
  {80}},\ \bibinfo {pages} {041309} (\bibinfo {year} {2009})}\BibitemShut
  {NoStop}%
\bibitem [{\citenamefont {Haupt}\ \emph {et~al.}(2009)\citenamefont {Haupt},
  \citenamefont {Novotn\'y},\ and\ \citenamefont {Belzig}}]{Haupt:2009}%
  \BibitemOpen
  \bibfield  {author} {\bibinfo {author} {\bibfnamefont {F.}~\bibnamefont
  {Haupt}}, \bibinfo {author} {\bibfnamefont {T.}~\bibnamefont {Novotn\'y}}, \
  and\ \bibinfo {author} {\bibfnamefont {W.}~\bibnamefont {Belzig}},\ }\href
  {\doibase 10.1103/PhysRevLett.103.136601} {\bibfield  {journal} {\bibinfo
  {journal} {Phys. Rev. Lett.}\ }\textbf {\bibinfo {volume} {103}},\ \bibinfo
  {pages} {136601} (\bibinfo {year} {2009})}\BibitemShut {NoStop}%
\bibitem [{\citenamefont {Novotn\'y}\ \emph {et~al.}(2011)\citenamefont
  {Novotn\'y}, \citenamefont {Haupt},\ and\ \citenamefont
  {Belzig}}]{Novotny:2011}%
  \BibitemOpen
  \bibfield  {author} {\bibinfo {author} {\bibfnamefont {T.}~\bibnamefont
  {Novotn\'y}}, \bibinfo {author} {\bibfnamefont {F.}~\bibnamefont {Haupt}}, \
  and\ \bibinfo {author} {\bibfnamefont {W.}~\bibnamefont {Belzig}},\ }\href
  {\doibase 10.1103/PhysRevB.84.113107} {\bibfield  {journal} {\bibinfo
  {journal} {Phys. Rev. B}\ }\textbf {\bibinfo {volume} {84}},\ \bibinfo
  {pages} {113107} (\bibinfo {year} {2011})}\BibitemShut {NoStop}%
\bibitem [{\citenamefont {Wheeler}\ \emph {et~al.}(2010)\citenamefont
  {Wheeler}, \citenamefont {Russom}, \citenamefont {Evans}, \citenamefont
  {King},\ and\ \citenamefont {Natelson}}]{Wheeler:2010}%
  \BibitemOpen
  \bibfield  {author} {\bibinfo {author} {\bibfnamefont {P.~J.}\ \bibnamefont
  {Wheeler}}, \bibinfo {author} {\bibfnamefont {J.~N.}\ \bibnamefont {Russom}},
  \bibinfo {author} {\bibfnamefont {K.}~\bibnamefont {Evans}}, \bibinfo
  {author} {\bibfnamefont {N.~S.}\ \bibnamefont {King}}, \ and\ \bibinfo
  {author} {\bibfnamefont {D.}~\bibnamefont {Natelson}},\ }\href {\doibase
  10.1021/nl904052r} {\bibfield  {journal} {\bibinfo  {journal} {Nano Letters}\
  }\textbf {\bibinfo {volume} {10}},\ \bibinfo {pages} {1287} (\bibinfo {year}
  {2010})},\ \bibinfo {note} \BibitemShut {NoStop}%
\bibitem [{\citenamefont {Chen}\ \emph {et~al.}(2012)\citenamefont {Chen},
  \citenamefont {Wheeler},\ and\ \citenamefont {Natelson}}]{Chen:2012}%
  \BibitemOpen
  \bibfield  {author} {\bibinfo {author} {\bibfnamefont {R.}~\bibnamefont
  {Chen}}, \bibinfo {author} {\bibfnamefont {P.~J.}\ \bibnamefont {Wheeler}}, \
  and\ \bibinfo {author} {\bibfnamefont {D.}~\bibnamefont {Natelson}},\ }\href
  {\doibase 10.1103/PhysRevB.85.235455} {\bibfield  {journal} {\bibinfo
  {journal} {Phys. Rev. B}\ }\textbf {\bibinfo {volume} {85}},\ \bibinfo
  {pages} {235455} (\bibinfo {year} {2012})}\BibitemShut {NoStop}%
\bibitem [{\citenamefont {Kumar}\ \emph {et~al.}(2012)\citenamefont {Kumar},
  \citenamefont {Avriller}, \citenamefont {Yeyati},\ and\ \citenamefont
  {van Ruitenbeek}}]{Kumar:2012}%
  \BibitemOpen
  \bibfield  {author} {\bibinfo {author} {\bibfnamefont {M.}~\bibnamefont
  {Kumar}}, \bibinfo {author} {\bibfnamefont {R.}~\bibnamefont {Avriller}},
  \bibinfo {author} {\bibfnamefont {A.~L.}\ \bibnamefont {Yeyati}}, \ and\
  \bibinfo {author} {\bibfnamefont {J.~M.}\ \bibnamefont {van Ruitenbeek}},\
  }\href {\doibase 10.1103/PhysRevLett.108.146602} {\bibfield  {journal}
  {\bibinfo  {journal} {Physical Review Letters}\ }\textbf {\bibinfo {volume}
  {108}},\ \bibinfo {pages} {146602} (\bibinfo {year} {2012})}\BibitemShut
  {NoStop}%
\bibitem [{\citenamefont {Djukic}\ \emph {et~al.}(2005)\citenamefont {Djukic},
  \citenamefont {Thygesen}, \citenamefont {Untiedt}, \citenamefont {Smit},
  \citenamefont {Jacobsen},\ and\ \citenamefont {van
  Ruitenbeek}}]{Djukic:2005}%
  \BibitemOpen
  \bibfield  {author} {\bibinfo {author} {\bibfnamefont {D.}~\bibnamefont
  {Djukic}}, \bibinfo {author} {\bibfnamefont {K.~S.}\ \bibnamefont
  {Thygesen}}, \bibinfo {author} {\bibfnamefont {C.}~\bibnamefont {Untiedt}},
  \bibinfo {author} {\bibfnamefont {R.~H.~M.}\ \bibnamefont {Smit}}, \bibinfo
  {author} {\bibfnamefont {K.~W.}\ \bibnamefont {Jacobsen}}, \ and\ \bibinfo
  {author} {\bibfnamefont {J.~M.}\ \bibnamefont {van Ruitenbeek}},\ }\href
  {\doibase 10.1103/PhysRevB.71.161402} {\bibfield  {journal} {\bibinfo
  {journal} {Phys. Rev. B}\ }\textbf {\bibinfo {volume} {71}},\ \bibinfo
  {pages} {161402} (\bibinfo {year} {2005})}\BibitemShut {NoStop}%
\bibitem [{\citenamefont {Wu}\ \emph {et~al.}(2007)\citenamefont {Wu},
  \citenamefont {Queipo}, \citenamefont {Nasibulin}, \citenamefont {Tsuneta},
  \citenamefont {Wang}, \citenamefont {Kauppinen},\ and\ \citenamefont
  {Hakonen}}]{Wu:2007}%
  \BibitemOpen
  \bibfield  {author} {\bibinfo {author} {\bibfnamefont {F.}~\bibnamefont
  {Wu}}, \bibinfo {author} {\bibfnamefont {P.}~\bibnamefont {Queipo}}, \bibinfo
  {author} {\bibfnamefont {A.}~\bibnamefont {Nasibulin}}, \bibinfo {author}
  {\bibfnamefont {T.}~\bibnamefont {Tsuneta}}, \bibinfo {author} {\bibfnamefont
  {T.~H.}\ \bibnamefont {Wang}}, \bibinfo {author} {\bibfnamefont
  {E.}~\bibnamefont {Kauppinen}}, \ and\ \bibinfo {author} {\bibfnamefont
  {P.~J.}\ \bibnamefont {Hakonen}},\ }\href {\doibase
  10.1103/PhysRevLett.99.156803} {\bibfield  {journal} {\bibinfo  {journal}
  {Physical Review Letters}\ }\textbf {\bibinfo {volume} {99}},\ \bibinfo
  {pages} {156803} (\bibinfo {year} {2007})}\BibitemShut {NoStop}%
\bibitem [{\citenamefont {Yao}\ \emph {et~al.}(2006)\citenamefont {Yao},
  \citenamefont {Chen}, \citenamefont {Di~Ventra},\ and\ \citenamefont
  {Yang}}]{Yao:2006}%
  \BibitemOpen
  \bibfield  {author} {\bibinfo {author} {\bibfnamefont {J.}~\bibnamefont
  {Yao}}, \bibinfo {author} {\bibfnamefont {Y.~C.}~\bibnamefont {Chen}}, \bibinfo
  {author} {\bibfnamefont {M.}~\bibnamefont {Di~Ventra}}, \ and\ \bibinfo
  {author} {\bibfnamefont {Z.~Q.}\ \bibnamefont {Yang}},\ }\href {\doibase
  10.1103/PhysRevB.73.233407} {\bibfield  {journal} {\bibinfo  {journal}
  {Physical Review B}\ }\textbf {\bibinfo {volume} {73}},\ \bibinfo {pages}
  {233407} (\bibinfo {year} {2006})}\BibitemShut {NoStop}%
\bibitem [{\citenamefont {Feng}\ \emph {et~al.}(1986)\citenamefont {Feng},
  \citenamefont {Lee},\ and\ \citenamefont {Stone}}]{Feng:1986}%
  \BibitemOpen
  \bibfield  {author} {\bibinfo {author} {\bibfnamefont {S.}~\bibnamefont
  {Feng}}, \bibinfo {author} {\bibfnamefont {P.~A.}\ \bibnamefont {Lee}}, \
  and\ \bibinfo {author} {\bibfnamefont {A.~D.}\ \bibnamefont {Stone}},\ }\href
  {\doibase 10.1103/PhysRevLett.56.1960} {\bibfield  {journal} {\bibinfo
  {journal} {Physical Review Letters}\ }\textbf {\bibinfo {volume} {56}},\
  \bibinfo {pages} {1960} (\bibinfo {year} {1986})}\BibitemShut {NoStop}%
\bibitem [{\citenamefont {{D'Agosta}}\ \emph {et~al.}(2006)\citenamefont
  {{D'Agosta}}, \citenamefont {Sai},\ and\ \citenamefont
  {Di~Ventra}}]{Dagosta:2006}%
  \BibitemOpen
  \bibfield  {author} {\bibinfo {author} {\bibfnamefont {R.}~\bibnamefont
  {{D'Agosta}}}, \bibinfo {author} {\bibfnamefont {N.}~\bibnamefont {Sai}}, \
  and\ \bibinfo {author} {\bibfnamefont {M.}~\bibnamefont {Di~Ventra}},\ }\href
  {\doibase 10.1021/nl062316w} {\bibfield  {journal} {\bibinfo  {journal} {Nano
  Lett.}\ }\textbf {\bibinfo {volume} {6}},\ \bibinfo {pages} {2935} (\bibinfo
  {year} {2006})}\BibitemShut {NoStop}%
\bibitem [{\citenamefont {Tsutsui}\ \emph {et~al.}(2012)\citenamefont
  {Tsutsui}, \citenamefont {Kawai},\ and\ \citenamefont
  {Taniguchi}}]{Tsutsui:2012}%
  \BibitemOpen
  \bibfield  {author} {\bibinfo {author} {\bibfnamefont {M.}~\bibnamefont
  {Tsutsui}}, \bibinfo {author} {\bibfnamefont {T.}~\bibnamefont {Kawai}}, \
  and\ \bibinfo {author} {\bibfnamefont {M.}~\bibnamefont {Taniguchi}},\ }\href
  {\doibase 10.1038/srep00217} {\bibfield  {journal} {\bibinfo  {journal}
  {Scientific Reports}\ }\textbf {\bibinfo {volume} {2}} (\bibinfo {year}
  {2012}),\ 10.1038/srep00217}\BibitemShut {NoStop}%
\bibitem [{\citenamefont {L{\"u}}\ \emph {et~al.}(2010)\citenamefont {L{\"u}},
  \citenamefont {Brandbyge},\ and\ \citenamefont {Hedeg{\aa}rd}}]{Lu:2010}%
  \BibitemOpen
  \bibfield  {author} {\bibinfo {author} {\bibfnamefont {J.-T.}\ \bibnamefont
  {L{\"u}}}, \bibinfo {author} {\bibfnamefont {M.}~\bibnamefont {Brandbyge}}, \
  and\ \bibinfo {author} {\bibfnamefont {P.}~\bibnamefont {Hedeg{\aa}rd}},\
  }\href {\doibase 10.1021/nl904233u} {\bibfield  {journal} {\bibinfo
  {journal} {Nano Letters}\ }\textbf {\bibinfo {volume} {10}},\ \bibinfo
  {pages} {1657} (\bibinfo {year} {2010})}. \BibitemShut {NoStop}%
\end{thebibliography}


\end{document}